\documentstyle[preprint,prl,aps,epsfig,floats]{revtex}
\begin{document}
\preprint{
\vbox{\halign{&##\hfil         \cr
        & DESY 00-185          \cr
        & hep-ph/0012244       \cr
        & December 2000        \cr
        }}}

\title{Polarization of $\mbox{\boldmath$\Upsilon(nS)$}$ at the Tevatron}

\author{Eric Braaten}
\address{Physics Department, Ohio State University, Columbus, OH 43210, USA}
\author{Jungil Lee}
\address{Deutsches Elektronen-Synchrotron DESY, 
	D-22603 Hamburg, Germany}

\maketitle
\begin{abstract}
The polarization of inclusive $\Upsilon(nS)$ at the Fermilab Tevatron 
is calculated within the nonrelativistic QCD factorization framework.  
We use a recent determination of the NRQCD matrix elements from 
fitting the CDF data on bottomonium production 
from Run IB of the Tevatron.
The result for the polarization of $\Upsilon(1S)$ 
integrated over the transverse momentum bin $8<p_T<20$ GeV
is consistent with a recent measurement by the CDF Collaboration.
The transverse polarization of $\Upsilon(1S)$ 
is predicted to increase steadily for $p_T$ greater than about 10 GeV. 
The $\Upsilon(2S)$ and $\Upsilon(3S)$ are predicted to have
significantly larger transverse polarizations than $\Upsilon(1S)$.

\end{abstract}
\vfill \eject

The production of heavy quarkonium provides an ideal testing ground for 
our understanding of the production mechanisms for heavy quarks
and the nonperturbative QCD effects that bind the heavy quark-antiquark 
pair into quarkonium.  
The purely leptonic decays of the $J^{PC}=1^{--}$ quarkonium states
allow very clean measurements of their cross sections.
One observable that is particularly sensitive to both the heavy-quark
production mechanisms and the QCD binding effects is the 
polarization of the $J^{PC}=1^{--}$ states, which can be measured from the
angular distribution of the leptons from their decays.

The NRQCD factorization approach provides a systematic framework 
for calculating the inclusive production rates 
of heavy quarkonium \cite{B-B-L}. 
It is based on organizing the inclusive production rate
into a double expansion in powers of $\alpha_s(m_Q)$, 
the running coupling constant at the scale of the heavy-quark mass, 
and of $v_Q$, the typical relative velocity of the heavy quark 
in the quarkonium rest frame.
It should therefore be more reliable for bottomonium
than for charmonium, because both of the expansion parameters 
$\alpha_s(m_Q)$ and $v_Q$ are smaller.
In the NRQCD factorization approach, 
the larger-than-expected cross sections observed at hadron colliders
are explained by introducing phenomenological parameters
that describe the probabilities for the formation of the quarkonium
state from color-octet heavy quark-antiquark pairs \cite{B-F}.
Once those parameters are determined, the cross sections 
for polarized states are predicted without any additional parameters.

The most dramatic prediction that has emerged from
the NRQCD framework is that in hadron colliders,
$J^{PC}=1^{--}$ quarkonium states should become increasingly 
transversely polarized as the transverse momentum $p_T$ 
increases \cite{Cho-Wise,Beneke-Rothstein}.
Quantitative predictions of the polarization of the charmonium states 
$J/\psi$ and $\psi(2S)$ at the Fermilab Tevatron indicate 
that there should be little polarization for $p_T$ around 5 GeV, 
but that the transverse polarization should increase steadily 
at larger $p_T$ \cite{Beneke-Kramer,Leibovich,B-K-L}.
The first measurements of the polarization of 
$J/\psi$ and $\psi(2S)$ by the CDF collaboration \cite{CDF-pol} 
have shown no evidence for this predicted increase.
However, the discrepancies with the
theoretical predictions are significant only in the highest $p_T$ bin,
so a definitive conclusion must await the higher statistics
measurements that will be possible in Run II of the Tevatron.
 
The CDF Collaboration has recently measured the polarization
of inclusive $\Upsilon(1S)$ in Run IB of the Tevatron \cite{CDF-upsIb}.
The results for the $p_T$ bins from 2 to 20 GeV and 
from 8 to 20 GeV are both consistent with 
no polarization. Since the cross section falls rapidly with $p_T$, 
this indicates that there is little if any polarization 
for $p_T$ below about 10 GeV.  
To determine whether this result is compatible with the NRQCD prediction,
we need a quantitative calculation of the 
polarization for {\it inclusive} $\Upsilon(1S)$ mesons.

The theoretical ingredients needed to calculate the polarization
of {\it direct} $\Upsilon(1S)$ mesons have been available for several
years \cite{Leibovich,B-K-V}.  
They were used by Beneke and Kr\"amer and by Leibovich to predict 
the polarization of prompt $\psi(2S)$ at the Tevatron 
\cite{Beneke-Kramer,Leibovich}.
The calculation of the polarization of {\it inclusive} $\Upsilon(nS)$
is complicated by the fact that the inclusive signal includes
$\Upsilon(1S)$ mesons that come from decays of higher bottomonium
states.  Decays of $\chi_{b}(1P)$, $\Upsilon(2S)$, and $\chi_{b}(2P)$
account for about 27\%, 11\%, and 11\% of the inclusive 
$\Upsilon(1S)$ signal, respectively  \cite{CDF-chib}.
The missing ingredients in the calculation of the polarization of
inclusive $\Upsilon(1S)$ were the cross sections for polarized $\chi_{bJ}$.
The necessary parton cross sections were recently calculated 
by Kniehl and Lee \cite{Kniehl-Lee} and used to predict the
polarization of prompt $J/\psi$ at the Tevatron \cite{B-K-L}.

In this paper, we present quantitative predictions for the polarization 
of inclusive $\Upsilon(1S)$, $\Upsilon(2S)$, and $\Upsilon(3S)$ 
at the Tevatron using the NRQCD factorization formalism.
We use NRQCD matrix elements for direct bottomonium production
which were recently determined by Braaten, Fleming, and Leibovich
from an analysis of data from Run IB at the Tevatron \cite{B-F-L}.
Our result for the polarization of $\Upsilon(1S)$ is consistent
with the recent measurement by the CDF Collaboration
in the $p_T$ bin from 8 to 20 GeV.
We predict that the transverse polarization of $\Upsilon(1S)$
should increase steadily for $p_T$ greater than about 10 GeV,
and that the $\Upsilon(2S)$ and $\Upsilon(3S)$
should be even more strongly transversely polarized.

Since the $\Upsilon(nS)$ is a spin-1 particle, the projection 
$\lambda$ of its spin along any quantization axis should be $-1$, 0, or $+1$.
The polarization of $\Upsilon(nS)$ can be measured from the angular 
distribution of the leptons from its leptonic decays.
The angular distribution of the positive lepton with respect to the
$\Upsilon(nS)$ momentum in the hadron center-of-momentum frame
is proportional to $1 + \alpha \cos^2\theta$,
which defines the polarization variable $\alpha$.
Taking the $\Upsilon(nS)$ momentum in this frame to be the 
spin quantization axis, we define the  longitudinally polarized 
$\Upsilon(nS)$ to be the $\lambda=0$ state and denote it by $\Upsilon_L(nS)$.
The polarization variable $\alpha$ is related to the fraction $\xi$ 
of $\Upsilon(nS)$ mesons
that are longitudinally polarized by 
$\alpha= (1-3\xi)/(1+\xi)$.
The longitudinal polarization fraction 
$\xi = \sigma_L/\sigma$ is the ratio
of the inclusive cross section for $\Upsilon_L(nS)$ 
to the inclusive cross section for $\Upsilon(nS)$ summed over spins.

The cross section $\sigma$ for inclusive $\Upsilon(nS)$
is the sum of the direct cross section 
and the direct cross sections for the higher bottomonium states 
$\Upsilon(mS)$ and $\chi_b(mP)$ weighted by the 
inclusive branching fractions $B_{H \to \Upsilon(nS)}$
for $H \to \Upsilon(nS) + X$:
\begin{equation}
\sigma[\Upsilon(nS)]_{\rm inc} \;=\;
\sigma[\Upsilon(nS)] 
+ \sum_H B_{H \to \Upsilon(nS)} \, \sigma[H] \;.
\label{sig-tot}
\end{equation}
For $\Upsilon(3S)$, we consider only the direct channel, neglecting 
any possible feeddown from higher states, such as $\chi_b(3P)$.
For $\Upsilon(2S)$, we take into account the direct channel and the 
feeddown from $\chi_b(2P)$ and $\Upsilon(3S)$.
For $\Upsilon(1S)$, we take into account the direct channel and the 
feeddown from $\chi_b(1P)$, $\Upsilon(2S)$, $\chi_b(2P)$, and $\Upsilon(3S)$.
The inclusive branching fractions $B_{H \to \Upsilon(nS)}$
are given in Table I of Ref.~\cite{B-F-L}.

The cross section $\sigma_L$ for inclusive $\Upsilon_L(nS)$
is the sum of the direct cross section for $\Upsilon_L(nS)$ and
the direct cross sections for the higher spin states
$\Upsilon(mS)_\lambda$ and $\chi_{bJ}(mP)_\lambda$
weighted by $B_{H \to \Upsilon(nS)}$ and by the conditional
probability $P_{H_\lambda \to \Upsilon_L(nS)}$ for $H_\lambda$
to decay into $\Upsilon_L(nS)$ given that it decays into $\Upsilon(nS)$:
\begin{equation}
\sigma_L[\Upsilon(nS)]_{\rm inc} \;=\;
\sigma[\Upsilon_L(nS)] 
+ \sum_{H,\lambda} B_{H \to \Upsilon(nS)} \, 
	P_{H_\lambda \to \Upsilon_L(nS)} \, \sigma[H_\lambda] \;,
\label{sig-L}
\end{equation}
The conditional probabilities are given in Table \ref{tab:prob}.
For $\chi_{bJ}(nP)_\lambda \to \Upsilon_L(nS)$,
$n=1,2$, they are given by simple angular-momentum factors
for the radiative transition.
For $\chi_{bJ}(2P)_\lambda \to \Upsilon_L(1S)$
and $\Upsilon(mS)_\lambda \to \Upsilon_L(nS)$,
we must average over the various decay paths weighted by their 
branching fractions.  The important steps in the decay paths 
are of 3 kinds. The observed hadronic transitions  
$\Upsilon(mS) \to \Upsilon(nS) + \pi \pi$ preserve the spin $\lambda$.
For the radiative transitions 
$\chi_{bJ}(2P)_\lambda \to \Upsilon_L(1S) + \gamma$ and
$\Upsilon(mS)_\lambda \to \chi_{bJ}(nP)_{\lambda'}+ \gamma$ ,
the probabilities for each spin state
are given by simple angular-momentum factors \cite{C-W-T}.
Taking the weighted average over the decay paths, we obtain
the results in Table \ref{tab:prob}.

\begin {table}
\begin {center}
\begin {tabular}{lc|c|c}
$H$             & $\lambda$ & $\Upsilon(2S)$ & $\Upsilon(1S)$ \\
\hline
$\Upsilon(3S)$  &    0      & 0.768 $\pm$ 0.022 & 0.853 $\pm$ 0.009 \\
                &  $\pm 1$  & 0.116 $\pm$ 0.011 & 0.073 $\pm$ 0.005 \\
\hline
$\chi_{b2}(2P)$ &    0      &    ${2\over3}$    & 0.654 $\pm$ 0.003 \\
                &  $\pm 1$  &    ${1\over2}$    & 0.494 $\pm$ 0.001 \\
		&  $\pm 2$  &         0         & 0.012 $\pm$ 0.003 \\ 
$\chi_{b1}(2P)$ &    0      &         0         & 0.013 $\pm$ 0.003 \\
                &  $\pm 1$  &    ${1\over2}$    & 0.494 $\pm$ 0.002 \\
$\chi_{b0}(2P)$ &    0      &    ${1\over3}$    &    ${1\over3}$    \\
\hline
$\Upsilon(2S)$  &    0      &                   & 0.941 $\pm$ 0.009 \\
                &  $\pm 1$  &                   & 0.029 $\pm$ 0.005 \\
\hline
$\chi_{b2}(1P)$ &    0      &                   &    ${2\over3}$    \\
                &  $\pm 1$  &                   &    ${1\over2}$    \\
		&  $\pm 2$  &                   &         0         \\ 
$\chi_{b1}(1P)$ &    0      &                   &         0         \\
                &  $\pm 1$  &                   &    ${1\over2}$    \\
$\chi_{b0}(1P)$ &    0      &                   &    ${1\over3}$    \\
\end {tabular}
\end {center}
\caption{ \label {tab:prob}
	Conditional probabilities $P_{H_\lambda \to \Upsilon_L(nS)}$
	for the bottomonium spin state $H_\lambda$ to decay into a 
	longitudinally polarized $\Upsilon(nS)$ given that it
	decays into $\Upsilon(nS)$.}
\end {table}

The polarization variable $\alpha$ for $\Upsilon(nS)$ has been expressed 
as a ratio of linear combinations of the direct cross sections for
$\Upsilon(nS)$ and higher bottomonium states.
The {\it NRQCD factorization formula} for the direct cross section for 
a bottomonium state $H$ of momentum $P$
and spin quantum number $\lambda$ has the schematic form
\begin{equation}
d \sigma[p \bar p \to H_\lambda(P) + X] \;=\;
\sum_n d \sigma[p \bar p \to b \bar b_n(P) + X] \; 
	\langle O^{H_\lambda(P)}_n \rangle,
\label{sig-fact}
\end{equation}
where the summation index $n$ extends over
all the color and angular momentum states of the $b\bar b$ pair.
The $b \bar b$ cross section can be expressed as 
\begin{equation}
d \sigma[p \bar p \to b \bar b_n(P) + X] \;=\;
f_{i/p} \otimes f_{j/\bar p}
\otimes d \hat \sigma[ij \to b \bar b_n(P)+X],
\label{sig-fusion}
\end{equation}
where $f_{i/p}(x,\mu)$ and $f_{j/\bar p}(x,\mu)$ are parton distribution
functions (PDF's) and a sum over the partons $i,j$ is implied.
The parton cross sections $d \hat \sigma$
can be calculated using perturbative QCD.
All dependence on the state $H$ is contained within the
nonperturbative matrix elements $\langle O^{H_\lambda(P)}_n \rangle$.
In general, they are Lorentz tensors that depend on the momentum $P$ 
and the polarization tensor of $H_\lambda$.
The Lorentz indices
are contracted with those of $d \sigma$
to give a scalar cross section.
The symmetries of NRQCD can be used to reduce 
the tensor matrix elements $\langle O^{H_\lambda(P)}_n \rangle$ 
to scalar matrix elements  $\langle O^H_n \rangle$ that are
independent of $P$ and $\lambda$.
This reduces the variable $\alpha$ to a ratio of linear combinations
of the NRQCD matrix elements.

A nonperturbative analysis of NRQCD reveals
how the various matrix elements scale with the typical relative 
velocity $v$ of the heavy quarks.  
The most important matrix elements for the production of
the S-wave states $\Upsilon(nS)$ and $\eta_b(nS)$
can be reduced to one color-singlet parameter
$\langle O^{\Upsilon(nS)}_1(^3S_1) \rangle$, which scales like $v^3$,
and three color-octet parameters 
$\langle O^{\Upsilon(nS)}_8(^3S_1) \rangle$,
$\langle O^{\Upsilon(nS)}_8(^1S_0) \rangle$, and
$\langle O^{\Upsilon(nS)}_8(^3P_0) \rangle$,
all of which scale like $v^7$.
The most important matrix elements for the production of
the P-wave states $\chi_{bJ}(nP)$ and $h(nP)$ 
can be reduced to a color-singlet parameter
$\langle O^{\chi_{b0}}_1(^3P_0) \rangle$ 
and a single color-octet parameter 
$\langle O^{\chi_{b0}}_8(^3S_1) \rangle$, both of which scale like $v^5$.
At higher orders in $v$, so many new matrix elements enter 
that the predictive power of the NRQCD approach is lost.  
We therefore assume the matrix elements enumerated above 
are sufficient to describe the bottomonium cross sections.

The first determination of the color-octet matrix elements
for bottomonium production was a 
pioneering analysis by Cho and Leibovich \cite{Cho-Leibovich}
of the data on bottomonium production from Run IA of 
the Tevatron \cite{CDF-upsIa}.  Due to the limited statistics,
they had to use educated guesses for some of the matrix elements. 
An updated theoretical analysis based on the new CDF data from Run IB 
\cite{CDF-upsIb} 
has been made by Braaten, Fleming, and Leibovich \cite{B-F-L}.
Their color-singlet matrix elements 
are given in Table II of Ref.~\cite{B-F-L}.  
Those for $\Upsilon(nS)$ were determined 
phenomenologically from the leptonic decay rates of $\Upsilon(nS)$,
while those for $\chi_{bJ}(nP)$ were estimated from potential models.  
Their color-octet matrix elements 
are given in Table V of Ref.~\cite{B-F-L}.  
They were determined by fitting the CDF data 
on the differential cross sections for $\Upsilon(1S)$, $\Upsilon(2S)$, 
and $\Upsilon(3S)$ at $p_T >8$ GeV 
and from the fractions of $\Upsilon(1S)$ 
from the decays of $\chi_b(1P)$ and $\chi_b(2P)$ \cite{CDF-chib}.

The leading terms in the parton cross sections 
in (\ref{sig-fusion}) depend on the region of the  
transverse momentum $p_T$ of the $b \bar b$ pair.
For $p_T$ in the range 8 GeV $< p_T <$ 30 GeV, 
the leading terms are {\it fusion} contributions
from the parton processes $i j \to b \bar b + k$.
For $p_T$ much greater than $2 m_b$, {\it fragmentation} contributions from
parton processes such as $i j \to g + k$, followed by $g \to b \bar b$,
become important.  For charmonium, 
fragmentation effects change the differential cross sections
by less than 3\% at $p_T = 5$ and less than 11\% at $p_T = 10$ GeV.
Since $m_b$ is 3 times larger than $m_c$, we expect fragmentation effects 
to change the differential cross sections for bottomonium 
by less than 11\% for $p_T$ less than 30 GeV.
We will therefore avoid the complications of fragmentation 
by restricting our predictions to $p_T < 30$ GeV.

For $p_T$ much smaller than $2 m_b$, 
parton processes such as $i j \to b \bar b + ggg...$ 
involving the multiple emission of soft gluons
become important and it is necessary to resum their effects.  
In the analysis of Ref.~\cite{B-F-L}, this problem was avoided by 
using only the CDF data for $p_T > 8$ GeV to fit the 
color-octet matrix elements.  Since we will be using the 
matrix elements from that analysis, we will also restrict our 
predictions to $p_T > 8$ GeV.
 
Having restricting our attention to the region 8 GeV $< p_T <$ 30 GeV,
it should be safe to use only the fusion cross sections 
for $d \hat \sigma$ in (\ref{sig-fusion}).
We include the parton processes $i j \to b \bar b + k$,
with $i,j,k=g,q,\bar q$ and $q = u,d,s,c$.
We treat the $c$ quark as a massless parton.
The leading-order parton cross sections $d \hat \sigma$
are proportional to $\alpha_s^3(\mu)$.
They are given in Refs. \cite{Leibovich} 
and \cite{B-K-V} for all the relevant $b \bar b$ color and spin states
with the exception of color-singlet $^3P_J$ states,
for which they are given in 
Ref.~\cite{Kniehl-Lee}\footnote{
	There is a typographical error in Eq.(22) of Ref.~\cite{Kniehl-Lee}. 
	The color-octet matrix elements 
	$\langle {\cal O}^{\chi_{c0}}(^3S_1^{(8)}) \rangle
		= \langle {\cal O}_8^{\chi_{c0}}(^3S_1) \rangle$ 
	should be replaced by 
	$(2J+1) \langle {\cal O}^{\chi_{c0}}(^3S_1^{(8)}) \rangle$.}.

We follow the analysis of Ref.~\cite{B-F-L} as closely as possible.
We use a common renormalization and factorization scale $\mu$ for
$f_{i/p}$, $f_{j/\bar p}$, and $\alpha_s$,
taking its central value to be $\mu_T = (m_b^2 + p_T^2)^{1/2}$ 
and allowing it to vary within the range ${1 \over 2}\mu_T$ to $2\mu_T$.
We set $m_b=4.77 \pm 0.11$ GeV.
We consider two choices for the PDF's for comparison: 
CTEQ5L and MRST98LO \cite{PDF}.
We evaluate $\alpha_s(\mu)$ from the one-loop formula with $n_f=5$ 
using the boundary conditions $\alpha_s(M_Z)=0.127$ for CTEQ5L and 
$\alpha_s(M_Z)=0.125$ for MRST98LO. 
The cross section for direct $\Upsilon(nS)$ depends on the linear combination
$\langle O_8(^1S_0) \rangle + r \langle O_8(^3P_0) \rangle/m_b^2$,
with $r$ varying from 4.6 at $p_T = 8$ GeV to 3.2 at $p_T = 30$ GeV.
In the cross section for direct $\Upsilon_L(nS)$, 
$r$ varies from 7.6 at $p_T = 8$ GeV to 3.6 at $p_T = 30$ GeV.
We consider two extreme cases: $\langle O_8(^1S_0) \rangle=0$ and
$\langle O_8(^3P_0) \rangle=0$.
The NRQCD matrix elements and their statistical errors 
are given in Table II and V of Ref.~\cite{B-F-L}.  
The color-octet matrix elements in Table V also have an additional
upper and lower error associated with changing $\mu$ by a factor of 2.
This allows the correlation between the errors in $\mu$ 
and the color-octet matrix elements to be taken into account.

The polarization variable $\alpha$ can be expressed as a ratio of 
linear combinations of NRQCD matrix elements.
The errors in the matrix elements from Ref.~\cite{B-F-L} 
give large uncertainties in the numerator and the
denominator, but they tend to cancel in the ratio.
As our central value for $\alpha$, we take the average value
from the 4 combinations corresponding to the two
choices $\langle O_8(^1S_0) \rangle=0$ or $\langle O_8(^3P_0) \rangle=0$ 
and the two choices of PDF's.
The deviations among the 4 combinations are included in our error.
We compute the errors from the remaining parameters by varying each one
independently for the choices MRST98LO and $\langle O_8(^3P_0) \rangle=0$
and using the central values of the other parameters.
The parameters include the scale $\mu$, the mass $m_b$, 
5 color-singlet matrix elements \cite{B-F-L}, 
8 color-octet matrix elements \cite{B-F-L},
and 18 exclusive branching fractions for quarkonium transitions \cite{PDG}.  
To compute the variation from $\mu$, 
we change $\mu$ by a factor of 2 or $1 \over 2$ in the PDF's and in
$\alpha_s$ and simultaneously shift all the matrix elements 
by the second upper or lower error in Table V of Ref. \cite{B-F-L}.
This takes into account the large correlation between the choice of $\mu$ and
the values of the color-octet matrix elements.
All the variations are added in quadrature to obtain the final
error on $\alpha$.

\begin{figure}
\begin{center}
\epsfig{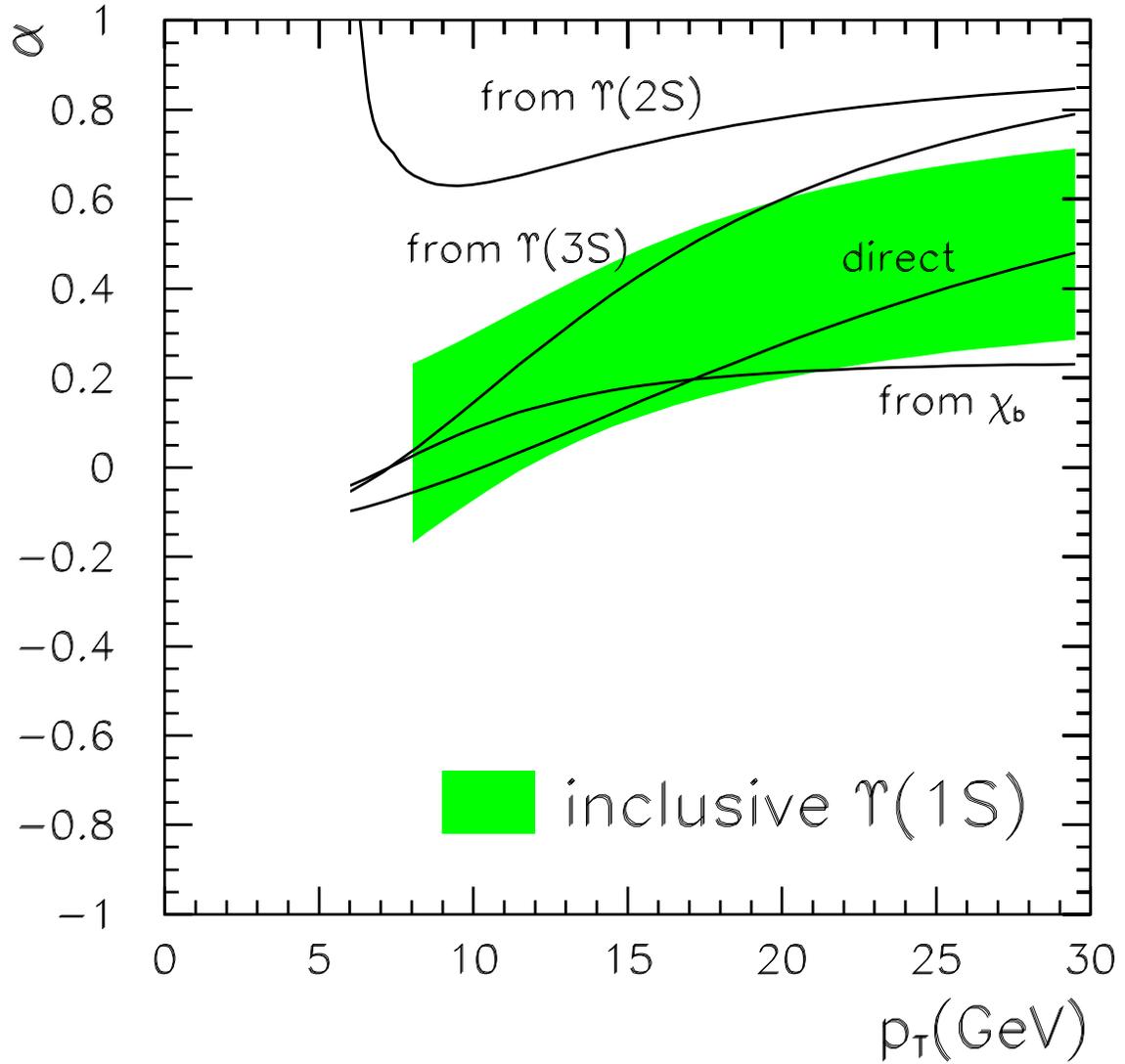}
\end{center}
\caption{
Polarization variable $\alpha$ vs.\ $p_T$ at $\sqrt{s}=2.0$ TeV
for inclusive $\Upsilon(1S)$ (shaded band).
The curves are the central values for direct $\Upsilon(1S)$, 
$\Upsilon(1S)$ from $\Upsilon(2S)+\pi \pi$, and
$\Upsilon(1S)$ from $\chi_b(1P)+\gamma$ or $\chi_b(2P)+\gamma$.
}
\label{fig-Ups1}
\end{figure}

\begin{figure}
\begin{center}
\epsfig{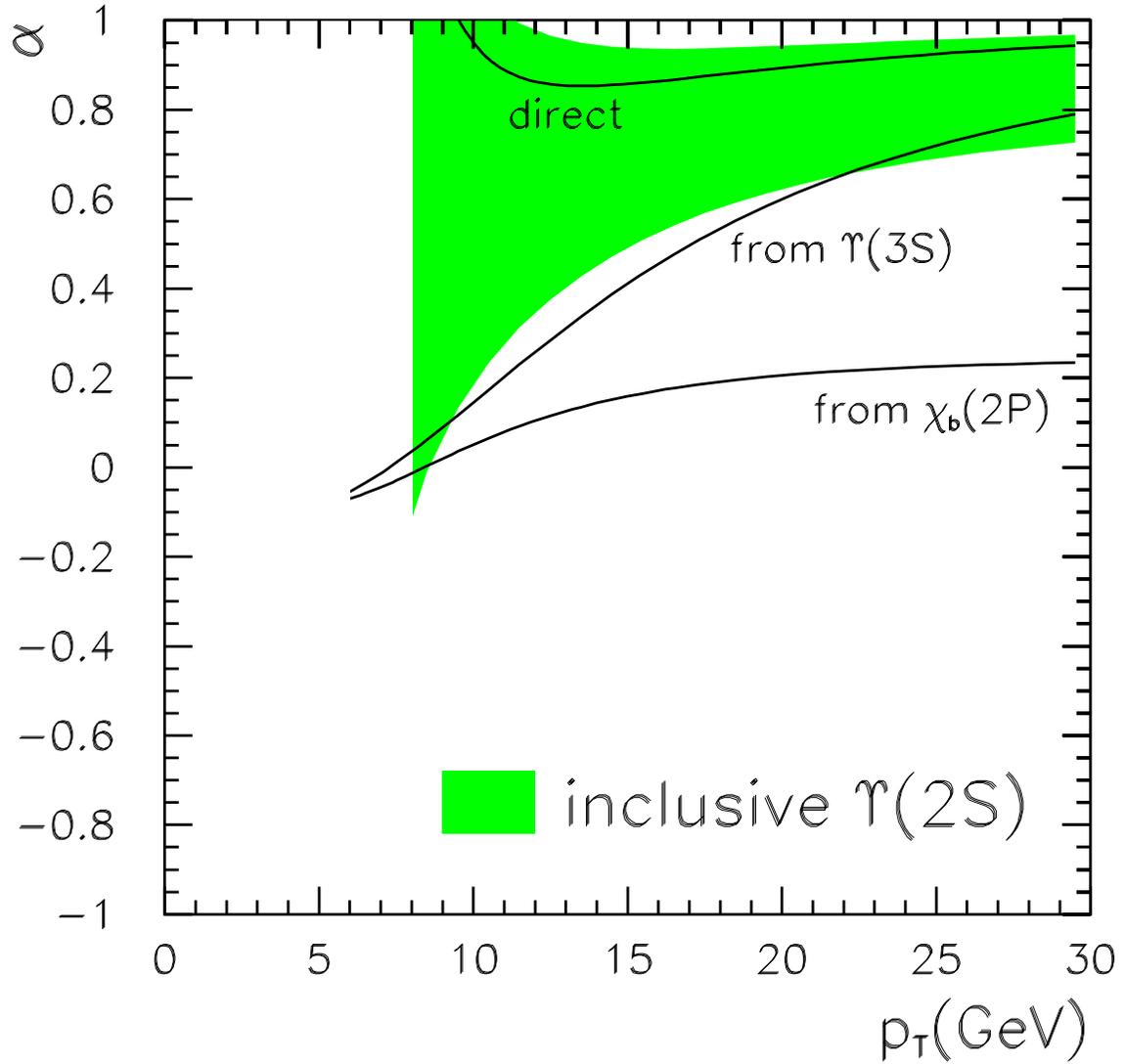}
\end{center}
\caption{
Polarization variable $\alpha$ vs.\ $p_T$ at $\sqrt{s}=2.0$ TeV
for inclusive $\Upsilon(2S)$ (shaded band).
The curves are the central values for direct $\Upsilon(2S)$, 
$\Upsilon(2S)$ from $\Upsilon(3S)+\pi \pi$, and
$\Upsilon(2S)$ from $\chi_b(2P)+\gamma$.
}
\label{fig-Ups2}
\end{figure}

\begin{figure}
\begin{center}
\epsfig{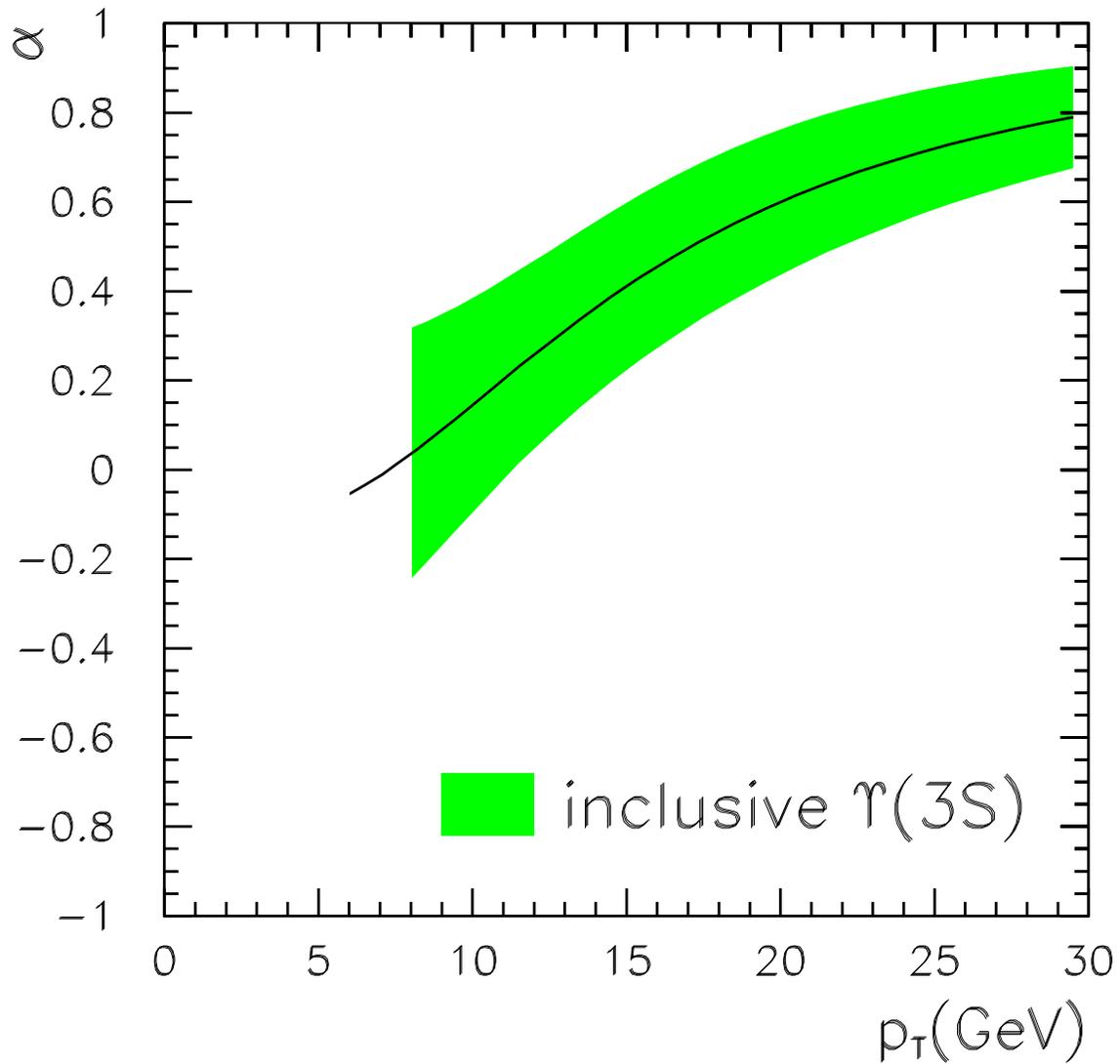}
\end{center}
\caption{
Polarization variable $\alpha$ vs.\ $p_T$ at $\sqrt{s}=2.0$ TeV
for $\Upsilon(3S)$ (shaded band).
The curve is the central value.
}
\label{fig-Ups3}
\end{figure}

Our results for the polarization variable $\alpha$ 
for $\Upsilon(1S)$, $\Upsilon(2S)$, and $\Upsilon(3S)$ 
for the Tevatron at $\sqrt{s} =$ 2.0 TeV are shown as shaded bands 
in Figs.~\ref{fig-Ups1}, \ref{fig-Ups2}, and \ref{fig-Ups3}, respectively.
The curves in Figs.~\ref{fig-Ups1}--\ref{fig-Ups3} 
are the central values of $\alpha$ for direct $\Upsilon(nS)$, 
for $\Upsilon(nS)$ from $\chi_b(mP) + \gamma$, 
and for $\Upsilon(nS)$ from $\Upsilon(mS) + \pi \pi$.
These channels together provide a complete decomposition of the inclusive
rate.  The fractions of the inclusive rate from each of these channels 
vary slowly with $p_T$ and add up to 1.  
For inclusive $\Upsilon(1S)$, the fractions have been measured by the CDF 
collaboration to be approximately 52\% for direct $\Upsilon(1S)$
and 27\%, 11\%, and 11\% for $\Upsilon(1S)$ from $\chi_b(1P)$, 
$\Upsilon(2S)$, and $\chi_b(2P)$, respectively \cite{CDF-chib}. 
For inclusive $\Upsilon(2S)$, the fractions can be estimated 
from the analysis of Ref.~\cite{B-F-L}
to be approximately 73\% for direct $\Upsilon(2S)$
and 25\% and  
2\% for $\Upsilon(2S)$ from $\chi_b(2P)$ and $\Upsilon(3S)$,
respectively.  For $\Upsilon(3S)$, we ignored any possible
feeddown from higher states.

The predictions for $\alpha$ for $\sqrt{s} =$ 1.8 TeV 
are essentially identical to the predictions for $\sqrt{s} =$ 2.0 TeV
in Figs.~\ref{fig-Ups1}--\ref{fig-Ups3}.
The cross sections $\sigma_L$ and $\sigma$ 
are both smaller by about 16\%, but the change cancels in the ratio.
Integrating over $p_T$ in the range 8 GeV $< p_T < 20$ GeV,
we obtain $\alpha = 0.13 \pm 0.18$.
This is in good agreement with
the value measured by the CDF collaboration:
$\alpha = 0.03 \pm 0.28$ \cite{CDF-upsIb}. 

In Figs.~\ref{fig-Ups1}--\ref{fig-Ups3}, the curves for $\alpha$ for direct
$\Upsilon(nS)$ and for $\Upsilon(nS)$ from $\Upsilon(mS)+ \pi \pi$
increase steadily with $p_T$.  The curves for 
$\Upsilon(nS)$ from $\chi_b(mP)+\gamma$ increase at first, but then flatten out 
at a value of $\alpha$ just above 0.2.
Thus the feeddown from $\chi_b(mP)$ tends to wash out the polarization.
The fraction of $\Upsilon(nS)$ from the $\chi_b$ states is 
approximately 38\% for $n=1$, 25\% for $n=2$, 
and probably smaller yet for $n=3$.
We therefore expect $\alpha$ to increase more rapidly with $p_T$
for $\Upsilon(2S)$ and $\Upsilon(3S)$ than for $\Upsilon(1S)$.
A much larger transverse polarization for $\Upsilon(2S+3S)$ 
than for $\Upsilon(1S)$ has also recently been observed in bottomonium 
production in p-Cu collisions at $\sqrt{s} =$ 38.8 GeV \cite{E866}. 

The predictions in Figs.~\ref{fig-Ups1}--\ref{fig-Ups3}
are based on NRQCD matrix elements extracted from CDF data in the range 
8 GeV $< p_T <$ 20 GeV.  The extrapolation of these results to lower 
values of $p_T$ should be considered unreliable. 
This is evident from the curves for direct $\Upsilon(1S)$ from $\Upsilon(2S)$
in Fig.~\ref{fig-Ups1} and direct $\Upsilon(2S)$ in Fig.~\ref{fig-Ups2}.
The dramatic changes in these curves at the smaller values of $p_T$ 
are artifacts of the 
fit of Ref.~\cite{B-F-L} having given negative central values 
for the color-octet matrix elements $\langle O_8(^1S_0) \rangle$
or $\langle O_8(^3P_0) \rangle$ for $\Upsilon(2S)$.

In Run II of the Tevatron, the much higher statistics  
will allow more accurate measurements of the polarization of 
$\Upsilon(1S)$ in several bins of $p_T$.
It may also allow measurements of the polarizations of 
$\Upsilon(2S)$ and $\Upsilon(3S)$.  
Significant improvements in the theoretical predictions are also possible.
The most important improvement is taking into account 
the effects of multiple soft-gluon
emission at low $p_T$.  This would allow more accurate determinations 
of the color-octet matrix elements, since the entire $p_T$ range 
of the CDF data from Run I could then be used in the fits.
It is also important to include fragmentation effects, 
so that the predictions can be extrapolated with confidence to large $p_T$.

In conclusion, we have calculated the polarization of the $\Upsilon(nS)$
states at the Tevatron using the NRQCD factorization approach.  
Our result is compatible with the recent CDF measurement 
for $\Upsilon(1S)$ in the $p_T$ bin from
8 to 20 GeV, which is consistent with no polarization.
However, our results also indicate that a nonzero transverse polarization 
should be observable for all three $\Upsilon(nS)$ states
in Run II of the Tevatron.

J.L. thanks Sungwon Lee and Heuijin Lim for their useful advice.
This work was supported in part by the US Department of Energy Division of 
High Energy Physics under grant DE-FG02-91-ER40690
and by the KOSEF and the DFG through
the German-Korean scientific exchange program DFG-446-KOR-113/137/0-1.


\newpage

\begin{references}

\bibitem{B-B-L}
G. T. Bodwin, E. Braaten, and G. P. Lepage,
Phys.\ Rev.\ D {\bf 51}, 1125 (1995); {\bf 55}, 5855(E) (1997).

\bibitem{B-F}
E. Braaten and S. Fleming,
Phys.\ Rev.\ Lett.\ {\bf 74}, 3327 (1995).

\bibitem{Cho-Wise}
P. Cho and M. B. Wise,
Phys.\ Lett.\ B {\bf 346}, 129 (1995).

\bibitem{Beneke-Rothstein}
M. Beneke and I. Z. Rothstein,
Phys.\ Lett.\ B {\bf 372}, 157 (1996); {\bf 389}, 769(E) (1996).

\bibitem{Beneke-Kramer}
M. Beneke and M. Kr\"amer,
Phys.\ Rev.\ D {\bf 55}, 5269 (1997).

\bibitem{Leibovich}
A. K. Leibovich,
Phys.\ Rev.\ D {\bf 56}, 4412 (1997).

\bibitem{B-K-L}
E. Braaten, B. Kniehl and J. Lee,
Phys. Rev. D {\bf 62}, 094005 (2000).

\bibitem{CDF-pol} 
CDF Collaboration, T. Affolder {\it et al.},
Phys. Rev. Lett. {\bf 85}, 2886 (2000). 

\bibitem{CDF-upsIb} 
R. Cropp (CDF Collaboration), hep-ex/9910003;  
V. Papadimitriou, FERMILAB-Conf-00-308-E.

\bibitem{B-K-V}
M. Beneke, M. Kr\"amer, and M. V\"anttinen,
Phys.\ Rev.\ D {\bf 57}, 4258 (1998).

\bibitem{CDF-chib} 
T. Affolder {\it et al.} (CDF Collaboration), 
Phys. Rev. Lett. {\bf 84}, 2094 (2000). 

\bibitem{Kniehl-Lee}
B. Kniehl and J. Lee,
Phys. Rev. D {\bf 62}, 114027 (2000).


\bibitem{B-F-L}
E. Braaten, A. K. Leibovich, and S. Fleming,
hep-ph/0008091.


\bibitem{C-W-T}
P. Cho, M. B. Wise, and S. P. Trivedi,
Phys.\ Rev.\ D {\bf 51}, 2039 (1995).


\bibitem{Cho-Leibovich}
P. Cho and A. K. Leibovich, Phys. Rev. D {\bf 53}, 150 (1996);
        {\bf 53}, 6203 (1996).

\bibitem{CDF-upsIa} 

F. Abe et al. (CDF Collaboration), Phys. Rev. Lett. {\bf 75}, 4358 (1997).

\bibitem{PDF}
A. D. Martin, R. G. Roberts, W. J. Stirling, and R. S. Thorne,
Eur.\ Phys.\ J. C {\bf 4}, 463 (1998);
CTEQ Collaboration, H. L. Lai {\it et al.},
Eur.\ Phys.\ J. C {\bf 12}, 375 (2000).

\bibitem{PDG}
Particle Data Group, D. E. Groom {\it et al.},
Eur.\ Phys.\ J. C {\bf15}, 1 (2000).

\bibitem{E866}
C. N. Brown et al. (FNAL E866/NuSea Collaboration), hep-ex/0011030.

\end{references}
\end{document}